\documentclass{article}
\usepackage{spconf,amsmath,amssymb,multirow,graphicx,color,soul,CJKutf8,pgfplots,cite}

\newcommand{\beqn}{\begin{eqnarray}}
        \newcommand{\eeqn}{\end{eqnarray}}
\newcommand{\beqnx}{\begin{eqnarray*}}
        \newcommand{\eeqnx}{\end{eqnarray*}}

\DeclareMathOperator*{\argmaxB}{argmax}

\definecolor{highlighter_red}{RGB}{255, 191, 191}
\definecolor{highlighter_green}{RGB}{174, 242, 178}

\ninept
\pgfplotsset{compat = newest}

\pgfplotstableread{fig/encoder_recall.tsv}{\tablet}

\title{Optimizing Bilingual Neural Transducer with Synthetic Code-switching Text Generation}

\name{\hspace{-0.75cm}\begin{tabular}{c} Thien Nguyen*, Nathalie Tran*, Liuhui Deng, Thiago Fraga da Silva, Matthew Radzihovsky, Roger Hsiao,\\
  Henry Mason, Stefan Braun, Erik McDermott, Dogan Can, Pawel Swietojanski, Lyan Verwimp, Sibel Oyman,\\
  Tresi Arvizo, Honza Silovsky, Arnab Ghoshal, Mathieu Martel, Bharat Ram Ambati and Mohamed Ali\end{tabular}
\thanks{*: these authors contributed equally to this work.}
}

\address{Apple\\
{\small \tt \{t.n, nathalie\_tran, rhsiao\}@apple.com}}

\begin{document}

\maketitle

\begin{abstract}
   Code-switching describes the practice of using more than one language in the same sentence. 
In this study, we investigate how to optimize a neural transducer based bilingual automatic speech recognition (ASR) 
model for code-switching speech.
Focusing on the scenario where the ASR model is trained without supervised code-switching data,
we found that semi-supervised training and synthetic code-switched data
can improve the bilingual ASR system on code-switching speech.
We analyze how each of the neural transducer's encoders contributes towards code-switching
performance by measuring encoder-specific recall values, and
evaluate our English/Mandarin system on the ASCEND data set. Our final
system achieves 25\% mixed error rate (MER) on the ASCEND English/Mandarin code-switching test set
-- reducing the MER by 2.1\% absolute compared to the previous literature -- 
while maintaining good accuracy on the monolingual test sets.

\end{abstract}

\section{Introduction}
\label{sect:intro}

Multilingual automatic speech recognition (ASR) is challenging. In addition to supporting multiple languages,
the system should also be able to handle the phenomenon of code-switching (CS), in which multiple languages are used in the same conversation.
CS can occur between sentences (inter-sentential), as well as within the same sentence (intra-sentential).
Building an ASR model with CS capabilities remains difficult due to the confusion of phonesets among the languages, the influence of the speaker's
native language during pronunciation, and the sparsity of data resources~\cite{dalmia2021}.

End-to-end (E2E) ASR models such as CTC~\cite{graves2006}, LAS~\cite{chan2016}, and neural transducers~\cite{graves2012, jaitly2016}
map an input sequence of acoustic features to an output sequence
of labels. These models have achieved good accuracy and performance, but they require large amount of training data.
In a low-resourced CS setting, data sparsity becomes an impediment for these models.
Despite this, there remains a growing interest in tackling CS for E2E ASR~\cite{kannan2019}.
Many of these efforts exercise different ideas, ranging from the integration of language identification (LID)~\cite{shan2019, zhu2020, qiu2020},
to leveraging multiple monolingual encoders~\cite{dalmia2021, song2022}.
To take advantage of large amounts of unlabeled data, semi-supervised learning (SSL) approaches have been popular in multilingual~\cite{voxpopuli2021, Conneau2021}
and E2E~\cite{li2019ssl, Yu2020} settings. In particular, semi-supervised training~\cite{Kahn2020, Synnaeve2020, lugosch2021}
has yielded significant performance improvements.

The move to E2E modeling led to a surge of interest concerning how to best leverage text-only data to further improve the performance of the new models.
Integrating external Neural Network Language Models (NNLM) into E2E ASR
through LM fusion~\cite{chorowski17_interspeech, kim2020improved, mcdermott2019, meng2021} has been shown to work well in monolingual scenarios.
NNLMs have also helped improve ASR performance on CS due to the greater availability of CS text data relative to speech data,
particularly through text synthesis in the form of grammar rules~\cite{rizvi2021, hu2020} or data driven models~\cite{winata2019, tarunesh2021}.
Furthermore, as better machine translation models (MT) become available, creating synthetic CS text becomes a viable option. 

In this work, we investigate how to improve a bilingual transducer model on CS speech without access to supervised CS training data.
We analyze the strengths and weaknesses of our models, and propose semi-supervised training and MT based synthetic data approach to
improve accuracy on CS speech.
This paper provides three contributions. First, we show that semi-supervised training can produce a performant bilingual neural
transducer despite only leveraging monolingual data.
Second, we describe a method for analyzing the bottleneck of a transducer model, particularly how said model reacts
internally to a CS query when it is trained with only monolingual data. Last, we apply the lessons learned from
this analysis and show that an external NNLM fine-tuned on synthetic CS data can be used to improve recognition accuracy on CS speech,
while maintaining good monolingual performance.

\section{Multilingual Modeling}
\label{sect:model}

\subsection{Multilingual neural transducer}
\label{sect:transducer}
The multilingual ASR model for this work is based on the neural transducer~\cite{graves2012, jaitly2016}. This model, $P_{\texttt{T}}$, consists of three components:
an acoustic encoder $E_{\texttt{A}}$, a label encoder $E_{\texttt{L}}$, and a joiner network $J$. $E_{\texttt{A}}$ processes a sequence of acoustic features
$\textbf{x} = \left( x_1 \cdots x_T \right)$ to generate a sequence of hidden representations.
$E_{\texttt{L}}$ takes as input a label history $\textbf{y} = \left( y_1 \cdots y_u \right)$ and generates a different hidden vector.
At time frame $t$ and label position $u$, $J$ (a 2-layer MLP with a $\tanh$ activation) takes the sum of the hidden vectors from $E_{\texttt{A}}$ and $E_{\texttt{L}}$, and computes the next output using:
\begin{equation} \label{eqn:transducer_top_level}
    \begin{split}
        P_{\texttt{T}}( y_{u+1}|\textbf{x}, t, & \;\textbf{y}_{1:u}) = \mbox{Softmax} \big( J(E_{\texttt{A}}(\textbf{x}, t) + E_{\texttt{L}}(\textbf{y}_{1:u}))\big).
    \end{split}
\end{equation}

In this work, the model output is the union of the subwords extracted from each language. Each subword list is derived from the training transcripts
of each language using BPE~\cite{sennrich2016neural}. Hence, this multilingual model is the same as the monolingual counterpart except for the output layer.
The advantage of this simple architecture is that the model can handle CS speech by emitting subwords of different languages within a query.
However, the challenge is to help the model acquire this capability despite limited access to CS training data.

\subsection{Semi-supervised learning}
We further explore the use of SSL, which has proven to be effective in improving E2E ASR performance~\cite{li2019ssl, Yu2020, Kahn2020, Synnaeve2020, lugosch2021}.
The training procedure consists of a two-stage approach. A model is first trained with unsupervised data, and then fine-tuned on supervised data.
In this work, the semi-supervised training procedure leverages randomly sampled and anonymized untranscribed speech data.
Auto transcripts are generated using a hybrid ASR model for pretraining.
During training, we explore different mixtures of English and Mandarin data: each batch of training data contains $a$\% of English and $(100-a)$\% of Mandarin semi-supervised data ($a$ ranging from 0 to 100).
Fine-tuning is done on a combined English and Mandarin supervised dataset, which is also randomly sampled and anonymized.

\subsection{Multilingual neural network language model}


To leverage the large amount of available text data relative to the amount of available speech data, we introduce an external NNLM
$P_{\texttt{LM}}$ via shallow fusion with subtraction of an Internal Language Model (ILM) $P_{\texttt{ILM}}$~\cite{kanda2017, mcdermott2019, variani2020, meng2021},
estimated from the transducer as described in~\cite{meng2021}.
The ASR scores of non-blank output symbols during beam search decoding are computed as shown in equation \ref{eqn:nnlm_fusion}:
\begin{equation} \label{eqn:nnlm_fusion}
    \begin{split}
        \textbf{y}^{*} = \argmaxB_\textbf{y} \space \text{log} & P_{\texttt{T}}( \textbf{y}|\textbf{x}) + \\
        & \lambda_{\texttt{LM}} \text{log} P_{\texttt{LM}}(\textbf{y}) - \lambda_{\texttt{ILM}} \text{log} P_{\texttt{ILM}}(\textbf{y}).
    \end{split}
\end{equation}
The NNLM in this work shares the same model architecture as the label encoder. The only difference is
that the label encoder allows emission of blank symbols, which are not modeled in the external LM.
To control the influence of this LM during decoding, weights of the external LM and the label encoder can be adjusted using hyper-parameters
$\lambda_{\texttt{LM}}$ and $\lambda_{\texttt{ILM}}$ respectively.

\subsection{Code switching text generation}
\label{subsection:cstg}


To improve the accuracy of CS speech despite insufficient data, we use machine translation (MT) to generate synthetic CS text as tuning data for the NNLM.
The purpose is to help the NNLM to perform better at switching points between languages, as without exposure to such CS data
the NNLM would only observe monolingual data, which is likely to be a bottleneck when it comes to CS accuracy.
To create synthetic text data, we explore the syntactic constraint approach described by Hu et al.~\cite{hu2020}, which leverages the
Equivalence Constraint Theory~\cite{poplack1981}:




\begin{enumerate}
    \item Perform MT on the source language to create a parallel corpus;
    \item Generate alignments between the source and target sequences using an alignment algorithm \cite{dyer2013};
    \item Generate Part-of-Speech (POS) tags for each source and target sequence in the parallel corpus;
    \item Generate synthetic text data. For each aligned token: find the token's alignment, and the corresponding target token it is aligned to. If the POS label of the target token matches the source token, then we perform substitution.
\end{enumerate}

We used our in-house MT model to perform translation to obtain synthetic CS data. SpaCy \cite{spacy2020} was used to
generate POS tags of the respective languages, and fast align \cite{dyer2013} was used to create alignments. This allowed us to
generate two different datasets, one with a single token substitution in place, and another with single token or phrase.
A phrase substitution occurs when a consecutive set of alignments occur after the original matching token from the source query.
We do not generate synthetic CS sentences with more than two code-switch points.
We also do not take into consideration the POS labels for the consecutive tokens, as the positional order of such tags within a phrase
for each language are likely to be different. With a parallel translation corpus of 63M sequences, we generated 93M CS
sentences with a single token substitution, and 500M sentences with a token or phrase substitution.

\section{Experimental Results}
\label{sect:expr}

\subsection{Data}

Our neural transducers are trained on a mixture of monolingual English and Mandarin supervised data, which is randomly sampled and anonymized.
The English training set consists of 14k hours of data while the Mandarin training set has 16k hours of data.
For SSL, we use unsupervised English and Mandarin datasets, with 300k hours for each language.
These unsupervised datasets are also randomly sampled and anonymized.
Auto transcripts are generated using the corresponding monolingual hybrid ASR systems as described in~\cite{Huang2020}.
For language modeling, we leverage an independent but similarly constructed text-only corpus containing around 3B queries for English,
and another 3B for Mandarin. In summary, our ASR models are trained on randomly sampled and anonymized monolingual data to begin with.

We evaluate our approach on randomly sampled and anonymized English and Mandarin test sets, and an English/Mandarin CS dataset known as ASCEND~\cite{lovenia2022ascend}.
The English and Mandarin test sets are monolingual and each contains around 60 hours of data.
The ASCEND corpus contains around 10 hours of audio recordings collected from native Mandarin speakers.
Within this, there are around four hours of CS speech, which we gather in one dataset called \textbf{mixed}.
This dataset was further cleaned through omitting unintelligible speech marked with an [UNK] token.
Table~\ref{ascend:stats} captures the dataset statistics.

\begin{table}[t!]
    \centering
    \begin{tabular}{|l|llll|l|}
        \hline
        \textbf{}     & \textbf{train} & \textbf{test} & \textbf{val} & \textbf{total} & \textbf{mixed} \\
        \hline
        N. Queries    & 9467           & 1279          & 1067         & 11813          & 3118           \\
        N. Hours      & 8.26           & 0.87          & 0.85         & 9.98           & 3.98           \\
        \hline
    \end{tabular}
    \caption{\label{ascend:stats} ASCEND dataset distributions.}
\end{table}

We found that the majority of the CS queries (80\%) in this dataset are Mandarin with English substitutions.
Additionally, 76\% of the dataset contains at most two code-switch points. Within this distribution, 55\% consisted of a single CS token.

\subsection{ASR models, inference and evaluation}

The transducer consists of a 12-layer conformer based acoustic encoder~\cite{gulati2020} and a 6-layer transformer based label encoder.
$E_{\texttt{A}}$ takes 80-dimension mel-filter bank features as input with 10ms frame rate. It then uses a convolutional layer to downsample the acoustic features
by a factor of six before feeding the features to the conformer blocks. In our experiments,
we replace the batch normalization layer in the conformer block with a layer normalization layer. Each conformer block uses causal convolution with a kernel width of 15.
For self-attention, it has eight heads and each head has a depth of 128 dimension. By using masking, we allow the first three self-attention blocks to look at
one future frame, so the resulting lookahead for $E_{\texttt{A}}$ is 180ms. The rest of the model is causal and
therefore, the model is capable of performing online recognition.
For the output layer, we use the BPE table based on the setup in \cite{deng2022} with 14725 tokens, which is the union of English subwords and Chinese characters.
The total number of parameters of the model is around 110M.
The Adam optimizer is used in both stages with lr=$0.0001$ for semi-supervised training, and lr=$0.00025$ for fine-tuning.
The initial learning rate is exponentially decayed every epoch with a base of 0.96.

We employ a transformer based NNLM with six self-attention blocks, totaling 29M parameters.
This NNLM uses the same BPE table as the transducer, and its architecture is largely the same as $E_{\texttt{L}}$, but has its own projection and softmax layer instead.
During inference, we perform LM fusion with ILM subtraction to combine the scores from the transducer and the language model.
We use interpolation weights $\lambda_{\texttt{LM}}$ ranging from 0.2 to 0.3, and $\lambda_{\texttt{ILM}}$ ranging from 0.1 to 0.3
based on tuning experiments against the ASCEND validation sets. We observe that a low $\lambda_{\texttt{LM}}$ performs best across the decoding experiments.

For evaluation, we adopt the mixed error rate (MER) metric from ASCEND~\cite{lovenia2022ascend}, which computes token level error rate. A token can be an English word
or a Chinese character. This hybrid metric helps us to evaluate the system's accuracy on CS speech.
For monolingual test sets, MER is the word error rate for English, and character error rate for Mandarin.
Table~\ref{results:baseline} contains the results for our baseline monolingual and bilingual models.
These models are trained on supervised data only.
We observe that despite being only trained on a mixture of monolingual supervised data, our baseline bilingual model can handle two languages and CS to some degree.
Compared to the corresponding monolingual systems, the bilingual system shows a slight degradation in accuracy by 0.4\% absolute in Mandarin and 0.3\% absolute in English.

\subsection{Results with semi-supervised learning}


\begin{table}[t!]
    \centering
    \begin{tabular}{|l|lll|}
        \hline
        \textbf{System} & \textbf{ZH}  & \textbf{EN}  & \textbf{Mixed} \\ 
        \hline
        Mandarin        & \textbf{6.6} & ---          & 38.0           \\ 
        English         & ---          & \textbf{4.7} & 88.2           \\ 
        Bilingual       & 7.0          & 5.0          & \textbf{31.5}  \\ 
        \hline
    \end{tabular}
    \caption{\label{results:baseline} MER of monolingual and bilingual systems trained on supervised data.}
\end{table}

SSL improves the accuracy of the bilingual system on both monolingual and CS test sets as shown in Table~\ref{results:ssl}.
Each row of Table~\ref{results:ssl} corresponds to a bilingual model pretrained using different ratios of English and Mandarin unsupervised data. After pretraining,
the model is then fine tuned using all supervised data.
As expected, we observe that pretraining on 100\% of English data shows improvements on English while degrading the accuracy on Mandarin.
The opposite observation can be made when pretraining solely on Mandarin data. The 50:50 ratio shows improvements over all the test sets.
This model outperforms the bilingual baseline by 0.6\% absolute on both Mandarin and English test sets and 2.1\% absolute on the mixed ASCEND test set.
A mixing ratio biased towards Mandarin (75:25) shows further improvement on the CS test set by 3.4\% absolute.
Despite 100:0 performing the best on the mixed dataset, we use the 75:25 model as the baseline for further experiments as 100:0 regresses severely
for monolingual English relative to the mixed gain.

\begin{table}[t!]
    \centering
    \begin{tabular}{|l|lll|}
        \hline
        \textbf{Data Ratio} & \textbf{ZH}  & \textbf{EN}  & \textbf{Mixed} \\ 
        \hline
        None (Baseline)     & 7.0          & 5.0          & 31.5           \\ 
        \hline
        0:100               & 9.5          & 4.6          & 31.9           \\ %
        25:75               & 6.8          & \textbf{4.3} & 29.7           \\ %
        50:50               & 6.4          & 4.4          & 29.4           \\ %
        75:25               & \textbf{6.2} & 4.9          & 28.1           \\ %
        100:0               & \textbf{6.2} & 6.0          & \textbf{27.9}  \\ %
        \hline
    \end{tabular}
    \caption{\label{results:ssl} MER with Semi-Supervised Learning against different ratios. Ratios are defined as Mandarin:English.}
\end{table}

\subsection{Results with NNLM Fusion}
\label{subsection:results_lmfusion}
\begin{table}[t!]
    \centering
    \begin{tabular}{|l|l|lll|}
        \hline
        \textbf{SSL} & \textbf{NNLM} & \textbf{ZH}  & \textbf{EN}  & \textbf{Mixed} \\ 
        \hline
        None         & None          & 7.0          & 5.0          & 31.5           \\ 
        75:25        & None          & 6.2          & 4.9          & 28.1           \\ 
        \hline
        75:25        & 25:75         & 6.3          & \textbf{4.6} & 27.8           \\ 
        75:25        & 50:50         & 6.2          & \textbf{4.6} & 27.8           \\ 
        75:25        & 75:25         & \textbf{6.1} & 4.7          & \textbf{27.7}  \\ 
        \hline
    \end{tabular}
    \caption{\label{results:lmfusion_ssl} MER of the different SSL + NNLM Fusion ratio configurations.}
    \vspace*{-3mm}
\end{table}

Table \ref{results:lmfusion_ssl} shows the combined results of NNLM using the different ratio configurations based on the best bilingual 
SSL system. Adding the NNLM to the best SSL system via fusion produces a moderate improvement.
The best system achieves 4.7\% MER on English test set, 6.1\% MER on Mandarin test set and 
27.7\% MER on ASCEND mixed CS set, but most importantly, no regressions in performance were observed.
Similar to the SSL experiment, the ratio of the data in semi-supervised training could bias the system toward a certain language.

\subsection{Analysis}

The results in Table~\ref{results:lmfusion_ssl} show that the model can handle two languages including CS speech to some extent.
However, CS speech is clearly more challenging to the bilingual system.
To better understand the strengths and weaknesses of this model, we use a variant of neural transducer, the ``simple joiner'' model~\cite{kuang2022}.
Compared to the standard neural transducer,
the outputs of both $E_{\texttt{A}}$ and $E_{\texttt{L}}$ are projected by a linear layer and the joiner network simply adds
the logits as shown in Equation~\ref{eqn:transducer_simple_joiner}:
\begin{equation} \label{eqn:transducer_simple_joiner}
    \begin{split}
        P_\texttt{T}(y_{u+1}|\textbf{x}, t, & \;\textbf{y}_{1:u}) = \\
        & \mbox{Softmax} \big( \mbox{Linear} ( E_{\texttt{A}}(\textbf{x}, t) ) + \mbox{Linear} (E_{\texttt{L}}(\textbf{y}_{1:u}))  \big).
    \end{split}
\end{equation}

The simple joiner model has been shown to be effective in first pass decoding. However, since $E_{\texttt{A}}$ and $E_{\texttt{L}}$
each have their own output layers, we can examine the tokens preferred by each encoder at different time or decoding steps. As a result, we can examine
the hypotheses proposed by each encoder in a CS query, and this helps us to analyze the model.

For this analysis, we first train a simple joiner model with the same supervised data. Then, we use the model to decode the CS test set.
For each query, we treat all the tokens in the minor language as keywords. We determine whether a language is minor
based on the number of tokens in a query. By doing so, we can measure the recall of these keywords for each encoder. This informs us about
how well each encoder in the model performs when it comes to CS speech. For $E_{\texttt{A}}$, we can examine the top $N$ tokens at every time step.
Similarly, we look at the top $N$ tokens from $E_{\texttt{L}}$ for every decoding step based on the context,
where the context here is derived from the reference.
To allow the decoder to switch languages, one would expect the keywords in the minor language to appear in the top $N$ entries in both the
acoustic and label encoders. Hence, by analyzing the recall of these keywords, we can infer whether $E_{\texttt{A}}$ or $E_{\texttt{L}}$ has
problems when it comes to CS.


\begin{figure}[htb]
    \centering
    \begin{tikzpicture}[every node/.style={font=\footnotesize}]
        \begin{axis}[
                xmin = 1, xmax = 10,
                ymin = 0, ymax = 0.8,
                xtick distance = 1,
                ytick distance = 0.25,
                xtick=data,
                nodes near coords,
                nodes near coords style={/pgf/number format/fixed, /pgf/number format/precision=2},
                nodes near coords align={vertical},
                ylabel=Recall,
                xlabel=Top N,
                grid = both,
                minor tick num = 1,
                major grid style = {lightgray},
                minor grid style = {lightgray!25},
                width = 0.48\textwidth,
                height = 0.30\textwidth,
                legend cell align = {left},
                legend pos = south east,
                legend style={nodes={scale=0.8, transform shape}},
            ]

            \addplot[blue, mark = *] table [x = {x}, y = {y0}] {\tablet};
            \addplot[red, mark = +] table [x = {x}, y = {y1}] {\tablet};
            \addplot[brown, mark = .] table [x ={x}, y = {y2}] {\tablet};

            \legend{
                $E_{\texttt{A}}$,
                $E_{\texttt{L}}$,
                NNLM w/ CS
            }

        \end{axis}
    \end{tikzpicture}
    \vspace*{-3mm}
    \caption{Recall of the acoustic and label encoders of the minor language in code-switching data.}
    \label{fig:encoder_recall}
\end{figure}
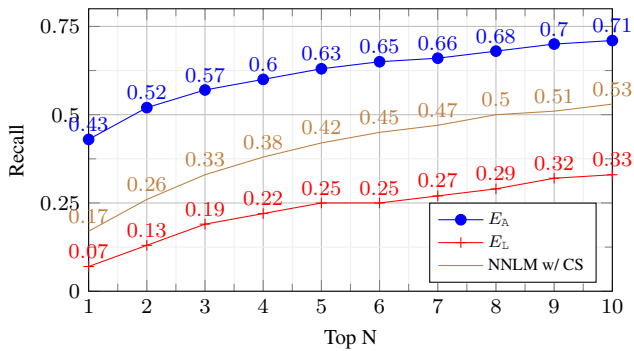

Figure~\ref{fig:encoder_recall} shows the recall of the acoustic and label encoders of the minor language at different top $N$ candidates. Larger $N$
implies higher recall as it considers more candidates, however, it also requires the decoder to use a higher beam to achieve the same recall.
This analysis gives us an interesting observation that while $E_{\texttt{A}}$ seems to have reasonable recall when it comes to the minor language,
$E_{\texttt{L}}$ is struggling at low 30\% recall. This means that even with the correct context, $E_{\texttt{L}}$ struggles to switch between languages.
This is understandable since the model is only trained on monolingual data. While $E_{\texttt{A}}$ might learn that some words in
different languages share similar acoustic characteristics, $E_{\texttt{L}}$ does not have the opportunity to make a similar discovery given that the
token sets are mostly mutually exclusive for each language and the transcripts are monolingual. In summary, if we want to improve the CS
accuracy for this model, we would need to figure out a way to encourage $E_{\texttt{L}}$ to switch languages when necessary.
Considering the lack of availability of code-switched speech data, we investigate if code-switched text-generation on an external language model
could be used to compensate for the shortcomings of $E_{\texttt{L}}$.

\subsection{Results on code-switching text generation}

\begin{table}[t!]
    \centering
    \begin{tabular}{|l|l|lll|}
        \hline
        \textbf{SSL} & \textbf{N. CS Samples}         & \textbf{ZH}  & \textbf{EN}  & \textbf{Mixed} \\ 
        \hline
        75:25        & None (Baseline)                & \textbf{6.1} & \textbf{4.7} & 27.7           \\ 
        \hline
        75:25        & 500m  (Token + Phrase)         & 6.3          & 5.3          & \textbf{26.5}  \\ 
        75:25        & 200m  (Token + Phrase)         & 6.3          & 5.3          & 26.7           \\ 
        75:25        & \textbf{93m  (Token + Phrase)} & 6.3          & 5.3          & \textbf{26.5}  \\ 
        \hline
        75:25        & 93m  (Token)                   & 6.6          & 5.6          & 26.7           \\ 
        75:25        & 37m  (Token)                   & 6.7          & 5.5          & 27.2           \\ 
        75:25        & 16m  (Token)                   & 6.9          & 5.4          & 26.8           \\ 
        \hline
    \end{tabular}
    \caption{\label{results:lmfusion_cs_aug_synt_con} MER of different SSL + NNLM models fine-tuned on synthetic code-switched data. Baseline numbers are from the best SSL + NNLM results.}
    \vspace*{-3mm}
\end{table}

The synthetically generated CS text is used for further fine-tuning the best NNLM in the previous experiment.
Table \ref{results:lmfusion_cs_aug_synt_con} shows the results of the syntactic constraint based approach.
We also investigate the relationship between the amount of synthetic data and accuracy.


Compared to our baseline (our best result from Table~\ref{results:lmfusion_ssl}), we observe a 1.2\% absolute improvement on ASCEND. 
Both the phrase and token level datasets improve on this baseline, but a single 
token based substitution incurs larger degradation on the monolingual tasks. This is likely caused 
by the model being less confused during the switch points, as a phrase is likely able to provide more context 
to determine when a language switch happens. Furthermore, the performance differences between the NNLMs fine-tuned 
on the synthetic data have statistical significance
against the baseline (with $p<0.001$). For reference, when evaluating the best model with the original ASCEND test set, we observe a MER of 25.0\%,
comparing favorably to the best reported result of 27.1\% MER in~\cite{lovenia2022ascend} despite not using any ASCEND data for training.
We also chart the recall of this NNLM against the CS queries in Figure~\ref{fig:encoder_recall}, finding a clear improvement from $E_{\texttt{L}}$.
This also provides a secondary perspective of the overall CS performance of the NNLM itself relative to the transducer encoders. 
Despite the improvement, we do not observe a strongly positive correlation in performance proportional to the amount of synthetic data. Our hypotheses include:
\begin{enumerate}
    \item It is possible that a fair portion of the generated dataset does not have any overlap with the distribution of the ASCEND dataset.
    \item Noisier data is likely to be present as more data is generated.
    \item Queries with multiple switch points were not generated.
    \item The quality of the synthetic data are subject to the performance of the MT/POS models, alignments, and the processes in between (e.g. standardized tokenisation and sanitization).
\end{enumerate}

\section{Conclusions}
\label{sect:conclude}

This paper presents three main contributions. First, we show that with semi-supervised training, we can develop a performant bilingual neural transducer 
with monolingual data. Second, we demonstrate how the simple joiner model can be used to analyze a neural transducer's behavior in general, and its CS
capabilities in particular. Last, we propose using synthetic data to further improve the model's CS accuracy
at the minor expense of some monolingual ASR accuracy. We believe this can be alleviated in future work by optimizing the fine-tuning procedure, improving the text generation process, 
or by adopting an LID approach to help the model determine whether the query is monolingual or bilingual.

\section{Acknowledgments}
\label{sect:acknowledge}
\vspace{-2mm}

We would like to thank Tatiana Likhomanenko, Takaaki Hori, Xinwei Li, and Xavier Suau Cuadros for their inputs and useful discussion.
\vspace{-2mm}

\bibliographystyle{IEEEbib}
\bibliography{bib/abbrev,bib/e2e}
\end{document}